\documentclass[showpacs,preprint,preprintnumbers,superscriptaddress,amsmath,amssymb,floatfix]{revtex4-1}

\usepackage{epsfig}
\usepackage{graphicx}
\usepackage{dcolumn}
\usepackage{bm}
\usepackage{times}
\usepackage{xcolor}
\usepackage[percent]{overpic}

\begin{document}


\title{Containing misinformation spreading in temporal social networks}

\author{Wei Wang}
\affiliation{Cybersecurity Research Institute, Sichuan University,
Chengdu 610065, China}

\author{Yuanhui Ma}
\affiliation{School of Mathematics, Southwest Jiaotong University,
  Chengdu 610031, China}

\author{Tao Wu}
\affiliation{School of Cyber Security and Information Law, Chongqing
  University of Posts and Telecommunications, Chongqing 400065, China}

\author{Yang Dai}\email{daiyang1980@gmail.com}
\affiliation{School of Economics and Management, Southwest Jiaotong
  University, Chengdu 610031, China}

\author{Xingshu Chen}\email{chenxsh@scu.edu.cn}
\affiliation{Cybersecurity Research Institute, Sichuan University,
Chengdu 610065, China}

\author{Lidia A. Braunstein}
\affiliation{ Instituto de Investigaciones F\'{i}sicas de Mar del
Plata (IFIMAR)-Departamento de F\'{i}sica,
Facultad de Ciencias Exactas y Naturales,
Universidad Nacional de Mar del Plata-CONICET,
Funes 3350, (7600) Mar del Plata, Argentina}
\affiliation{Center for Polymer Studies and Department of Physics,
Boston University, Boston, Massachusetts 02215, USA}

\date{\today}

\begin{abstract}

\noindent
Many researchers from a variety of fields including computer science,
network science and mathematics have focused on how to contain the
outbreaks of Internet misinformation that threaten social systems and
undermine societal health. Most research on this topic treats the
connections among individuals as static, but these connections change in
time, and thus social networks are also temporal networks.  Currently
there is no theoretical approach to the problem of containing
misinformation outbreaks in temporal networks. We thus propose a
misinformation spreading model for temporal networks and describe it
using a new theoretical approach. We propose a heuristic-containing (HC)
strategy based on optimizing final outbreak size that outperforms
simplified strategies such as those that are random-containing (RC) and
targeted-containing (TC). We verify the effectiveness of our HC strategy
on both artificial and real-world networks by performing extensive
numerical simulations and theoretical analyses. We find that the HC
strategy greatly increases the outbreak threshold and decreases the
final outbreak threshold.

\end{abstract}

\pacs{89.75.Hc, 87.19.X-, 87.23.Ge}

\maketitle

\section{Introduction} \label{sec:intro}

\noindent
Many communications platforms, e.g., Twitter, Facebook, email, WhatsApp,
and mobile phones, allow numerous ways of sharing information
\cite{vosoughi2018spread,del2016spreading,qiu2017limited,
  schmidt2017anatomy,castellano2009statistical,zhang2016dynamics}. One
task for researchers is developing ways to distinguish between true and
false information, i.e., between ``news'' and ``fake news''
\cite{kumar2018false}. This task is important because access to true
information is essential in the process of intelligent decision-making
\cite{funk2009spread,granell2013dynamical, wang2014asymmetrically}.  For
example, when the Severe Acute Respiratory Syndrome (SARS) spread across
Guangzhou, China in 2003, the {\it Chinese Southern Weekly\/} published
a newspaper article entitled ``There is a Fatal Flu in Guangzhou.''
This information was forwarded over 126 million times by TV news and in
other newspapers \cite{tai2007media,chow2003consensus}. Individuals
receiving this true information could adopt simple, effective protective
measures against being infected (e.g., by staying at home, washing
hands, or wearing masks). Misinformation, on the other hand, encourages
irrational behavior and reckless decision-making, and its spread can
undermine societal well-being and sway the outcome of elections
\cite{allcott2017social,soll2016long,
  bessi2015science,mocanu2015collective}. Bovet and Makse
\cite{bovet2018influence} analyzed 171 million tweets sent during the
five months prior to the 2016 US presidential election and found that
misinformation strongly affected the outcome of that election.

To contain the spread of misinformation we must understand the dynamic
information spreading mechanisms that facilitate it
\cite{pastor2015epidemic,de2018fundamentals,wang2018social,chen2018optimal,
  young2011dynamics,weiss2014adoption,iribarren2009impact,
  alvarez2019dynamic,di2018multiple,valdez2018role,yang2018competition}. Vosoughi et
al. \cite{vosoughi2018spread} examined true and fake information on
Twitter from 2006 to 2017 and found that misinformation spreads more
quickly than true information. Using the spreading mechanisms common in
real-data analysis, researchers have proposed several mathematical
models to describe the spreading dynamics of true and fake information
\cite{moreno2004dynamics,nekovee2007theory,
  zhou2007influence,isham2010stochastic,zhao2012sihr,
  kwon2013prominent}. Moreno et al. \cite{moreno2004dynamics} developed
mean-field equations to describe the spread of classical misinformation
on static scale-free networks that enables a theoretical study not
requiring extensive numerical simulations. Borge-Holthoefer and Moreno
\cite{borge2012absence} found that although there are no influential
spreaders in the classical misinformation model presented in
Ref.~\cite{moreno2004dynamics}, nodes with high $k$-cores and ranking
values are more likely to be the influential spreaders of true
information and also of infectious diseases
\cite{lu2016h,lu2016vital,kitsak2010identification,
  morone2015influence,pei2018theories}. When we include the burst
behavior of individuals in the misinformation model, hubs emerge as
influential nodes \cite{borge2013emergence}.  Using real-world data,
researchers found that social networks evolve with time, and thus
evolving temporal networks more accurately represent the topology of
real-world networks than static networks
\cite{holme2012temporal,holme2013temporal,
  holme2015modern,speidel2016temporal,masuda2013predicting,
  masuda2016guidance,li2017fundamental,nadini2018epidemic,
  liu2016social,kalantar2006survival,starnini2013modeling,speidel2016temporal}.

Researchers have found that the temporal nature of networks strongly
affect their spreading dynamics. Perra et al. \cite{perra2012activity}
found that in susceptible-infected-susceptible (SIS) epidemic spreading
a temporal network behavior suppresses the spreading more effectively
than a static integrated network. Researchers have also found that the
SIS and susceptible-infected-recovered (SIR) models on temporal networks
exhibit the same outbreak threshold
\cite{valdano2015analytical,valdano2018epidemic, onaga2017concurrency}.
Nadini et al. \cite{nadini2018epidemic} found that tightly connected
clusters in temporal networks inhibit SIR processes, but accelerate SIS
spreading. Pozzana et al. found that when node attractiveness---its role
as a preferential target for interactions---in temporal networks is
heterogeneous, the contagion process is altered
\cite{pozzana2017epidemic}.  Karsai et al. \cite{karsai2014time} found
that strong ties between individuals strongly inhibit the classical
spreading dynamics of misinformation.

Several strategies for containing the spread of misinformation in
temporal networks have been proposed
\cite{ogura2017optimal,liu2014controlling,
  lima2015disease,moinet2018effect}. Liu et
al. \cite{liu2014controlling} examined epidemic spreading on activity
driven temporal networks and developed mean-field based theoretical
approaches for three different control strategies, i.e., random,
targeted, and egocentric. The egocentric strategy is most effective. It
immunizes a randomly selected neighbor of a node in the observation
window. Other effective approaches using extensive numerical simulations
have been proposed
\cite{holme2014birth,masuda2013temporal,vestergaard2014memory}.  For
example, Holme and Liljeros \cite{holme2014birth} take into
consideration the time variation of nodes and edges and propose a
strategy for containing the outbreak of an epidemic based on the birth
and death of links.

Because there is still no theoretical approach to containing the spread
of misinformation in temporal networks, we here systematically examine
its spread in activity-driven networks. The rest of the paper is
organized as follows.  Section~\ref{model} describes the misinformation
spreading dynamics on temporal networks and develops a theory to
describe the spreading dynamics. Section~\ref{con} proposes three
containment strategies. Section~\ref{num} describes the results of our
extensive numerical stochastic simulations, which show that our
suggested theory agrees with the numerical simulations.
Section~\ref{con} presents our conclusions.

\section{Misinformation spreading in temporal networks}\label{model}

\noindent
We here introduce our model for the spreading dynamics of
misinformation in temporal networks.

\subsection{Mathematical descriptions for temporal networks}

\noindent
The widely used approaches mathematically describing temporal networks
tend to be either event-based or snapshot representations
\cite{masuda2016guidance}. The event-based representation approach
describes describes temporal networks using ordered events
$\{u_i,v_i,t_i,\Delta t_i; i=1,2,\cdots\}$, where node $u_i$ and $v_i$
are connected at time $t_i$ in the time period $\Delta t_i$. The
snapshot approach describes temporal networks using a discrete sequence
of static networks $\mathcal{G}=\{\mathcal{G}(1),\mathcal{G}(2),\cdots,
\mathcal{G}(t_{\rm max})\}$, where $\mathcal{G}(t)$ is the snapshot
network at time $t$, and $t_{\rm max}$ is the number of snapshots of the
temporal network.  Each snapshot network $\mathcal{G}(t)$, contains $N$
nodes, where $N$ is fixed, and $M_{t }$ edges. Thus the average temporal degree
of snapshot network $\mathcal{G}(t)$ is $\langle
k_{t}\rangle=2M_t/N$. Using the adjacency matrix, the temporal network
is $\mathcal{A}=\{\mathcal{A}(1),\mathcal{A}(2),\cdots,
\mathcal{A}(t_{\rm max})\}$, where $\mathcal{A}(t)$ is the adjacency
matrix of network $\mathcal{G}(t)$.

We here adopt the snapshot approach to describe temporal networks. As
the meaning of the adjacency matrix in static networks,
$\mathcal{A}_{uv}(t)=1$ when nodes $u$ and $v$ are connected at time
$t$, otherwise, $\mathcal{A}_{uv}(t)=0$. Thus the degree $k_{u}(t)$ of
node $u$ at time $t$ is $k_{u}(t)=\sum_{v=1}^N\mathcal{A}_{uv}(t)$ for
undirected temporal networks. The average degree of node $u$ in the
temporal network $\mathcal{G}$ is $\langle k_u(t)\rangle=\frac{1}{t_{\rm
    max}}\sum_{t=1}^ {t_{\rm max}}k_{u}(t)$. Knowing the adjacency
matrix {\bf $\mathcal{A}$} we obtain the eigenvalues of ${\bf
  \mathcal{A}}$. Here $\Lambda_1({\bf \mathcal{A})}$ $\Lambda_2({\bf
  \mathcal{A}})$, $\cdots$, and $\Lambda_N({\bf \mathcal{A}})$ are the
eigenvalues of ${\bf \mathcal{A}}$ in decreasing order. The spectral
radius is thus $\Lambda_1({\bf \mathcal{A}})$, which quantifies the
threshold outbreak of epidemics in temporal networks.

\subsection{Activity-driven networks}

\noindent
We use the classical activity-driven network \cite{perra2012activity} to
model a temporal network with $N$ nodes. We build the activity-driven
network using the following steps.

\begin{enumerate}

\item We assign to each node $i$ an activity potential value $x_i$
  according to a given probability density distribution $f(x)$. The
  activity of node $i$ is $a_i=\eta x_i$, where $\eta$ is a rescaling
  factor, i.e., at each time step, node $i$ is active with probability
  $a_i$. The higher the value of $\eta$, the higher the average degree
  of the temporal network. The higher the value of $a_i$, the higher the
  degree of node $i$. We assume that $f(x)$ follows a power-law
  function, i.e., $f(x)\sim x^{-\gamma}$, where $\gamma$ is the
  potential exponent. We make this assumption in order to generate a
  heterogeneous degree distribution temporal network. After further
  calculations we find
  $f(x)=\frac{\gamma-1}{\epsilon^{1-\gamma}}x^{-\gamma}$, $\langle
  k_t\rangle=2m\langle a\rangle$, and $\langle a\rangle=\eta\langle
  x\rangle$, where $\langle x\rangle=\int_\epsilon^\infty x f(x)dx$, and
  $\epsilon$ is the minimum value of the activity potential $x_i$.  Each
  active node has $m$ edges, and each edge randomly links to a network
  node. An edge connects the same pair of nodes with probability
  $m/N$. Note that in the thermodynamic limit of a sparse temporal
  network, there are no multiple edges between nodes and non-local loops

\item At the end of time step $t$, we delete all edges in network
  $\mathcal{G}(t)$.

\item We repeat steps (2) and (3) until $t_{\rm max}$ in order to
  generate temporal network $\mathcal{G}$.

\end{enumerate}

\subsection{Misinformation spreading model}

\begin{figure}
\begin{center}
\epsfig{file=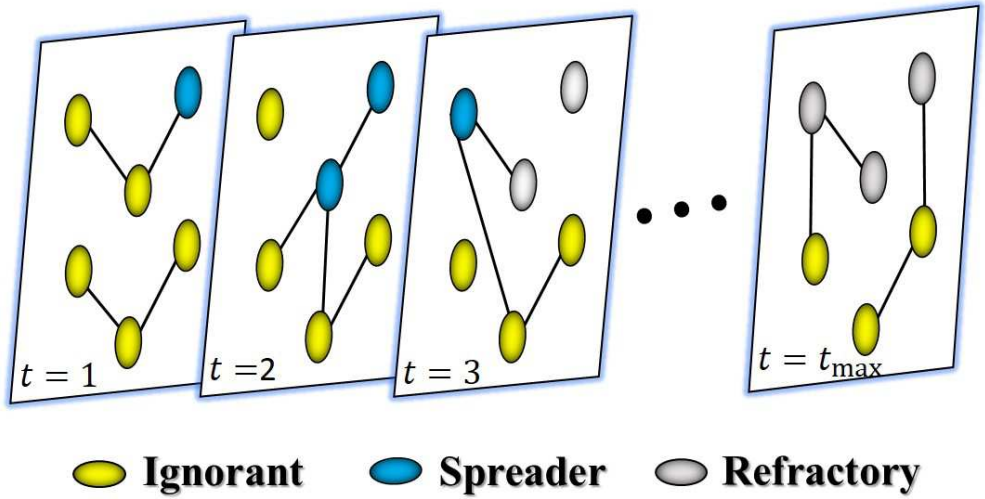,width=0.8\linewidth}
\caption{Illustration of misinformation spreading on temporal
  networks.}\label{illu}
\end{center}
\end{figure}

\noindent
We use an ignorant-spreader-refractory model to describe the
spreading dynamics of misinformation \cite{moreno2004dynamics}. Here
nodes are classified as either ignorant, spreader, or refractory.
Ignorant nodes are unaware that the information is false but are
susceptible to adopting it.  Spreader nodes are aware that the
information is false and are willing to transmit it to ignorant
nodes. Refractory nodes receive the misinformation but do not spread
it. The misinformation spreading dynamics on temporal networks evolves
as follows. We first randomly select a small fraction $\rho_0$ of
spreader nodes to be seeds in network $\mathcal{G}(t_0)$, where $1\leq
t_0\leq t_{\rm max}$. We designate the remaining $1-\rho_0$ nodes to be
ignorant.  At time step $t$ each spreader $i$ transmits with probability
$\lambda$ the misinformation to ignorant neighbors in network
$\mathcal{G}(t)$. In addition, each spreader $i$ becomes a refractory
node with a probability
\begin{equation} \label{rec_pro}
\mu_i(t)=1-(1-\mu)^{n(t)},
\end{equation}
where $\mu$ is the intrinsic recovery probability, and $n(t)$ is the
number of nodes in the spreader and refractory states of node $i$. The
dynamics evolve until there are no spreader nodes.  Note that when $t$
reaches $t_{\rm max}$, the misinformation spreads on $\mathcal{G}(1)$ in
the next time step. Figure~\ref{illu} shows misinformation spreading on
a temporal network.

\section{Theoretical analysis}

\noindent
We here develop a generalized discrete Markovian approach to describe
the misinformation spreading dynamics on temporal networks
\cite{valdano2015analytical}. We denote $I_i(t)$, $S_i(t)$, and $R_i(t)$
to be the fraction of nodes in the ignorant, spreader, and refractory
states, respectively, at time $t$. Because a node can only be in one of
the three states, $I_i(t)+ S_i(t)+R_i(t)=1$.

An ignorant node $i$, becomes a spreader with probability
$p_i\left(t\right)$ at time $t$, where
\begin{equation}\label{pit}
p_i\left(t\right) = 1-\prod_j{\left[1-\lambda \mathcal{A}_{ij}
\left(t\right) I_j\left(t\right) \right]}.
\end{equation}
Here $\prod_j{\left[1-\lambda \mathcal{A}_{ij} \left(t\right)
    I_j\left(t\right) \right]}$ is the probability that node $i$ has not
received any misinformation from neighbors at time $t$ in
$\mathcal{G}(t)$.  At time step $t+1$, node $i$ remains ignorant with a
probability
\begin{equation}\label{i}
I_i(t+1)=I_i(t)-I_i(t)p_i(t).
\end{equation}
The decrease of $I_i(t)$ is equal to the increase of $S_i(t)$, because
an ignorant node will become a spreader when it obtains the information
from neighbors in state $S$. In addition, spreader $i$ becomes
refractory with a probability $\mu_i(t)=1-(1-\mu)^{1+\sum_{j=1}^N
  \mathcal{A}_{ij}(t)[I_j(t)+R_j(t)]}$. Thus the evolution of node $i$
in the spreader state is
\begin{equation}\label{s}
S_i(t+1)=S_i(t)+I_i(t)p_i(t)-\mu_i(t)S_i(t).
\end{equation}
The evolution of node $i$ in the refractory state is
\begin{equation}\label{r}
R_i(t+1)=R_i(t)+\mu_i(t)S_i(t).
\end{equation}
Using Eqs.~(\ref{i})-(\ref{r}), we obtain the fraction of nodes at time
$t$ in each state,
\begin{equation}\label{each}
H(t)=\frac{1}{N}\sum_{i=1}^N H_i(t),
\end{equation}
where $H\in\{I,S,R\}$. Note that in the steady state there are no
spreader nodes, only refractory and stifler nodes. The fraction of nodes
that receive the misinformation in the final state is $R(\infty) =
R$. Here $R$ is the order parameter of a continuous phase transition with
$\lambda$. If the misinformation transmission probability $\lambda$ is
larger than the critical threshold, i.e., $\lambda>\lambda_c$, the size
of the global misinformation is of the order of the system size.
Otherwise, the global misinformation $R = 0$ for $\lambda\leq\lambda_c$
is in the thermodynamic limit.  At shorter times, a vanishingly small
fraction of nodes receive the misinformation, i.e.,
$S_i\left(t\right)\approx 0$ and
$R_i\left(t\right)=1-S_i\left(t\right)-I_i\left(t\right)\approx 0$.  The
recovery probability in Eq.~(\ref{rec_pro}) of a spreader node $i$ is
$\mu_i\left(t\right)\approx \mu$, since node $i$ must connect to a
spreader that supplies the misinformation, and there is a low
probability that it will connect to other spreader or refractory
neighbors. Thus Eq.~(\ref{s}) can be rewritten
\begin{equation}\label{s_line}
S_i(t+1)\approx\sum_{j=1}^N [(1-\mu)+\lambda\mathcal{A}_{ij}
(t)\delta_{ij}]S_j(t),
\end{equation}
where $\delta_{ij}$ is the Kronecker delta function, i.e.,
$\delta_{ij}=1$ if $i=j$, and zero otherwise. We define the transmission
tensor $\mathbf{M}$ to be
\begin{equation}\label{tensor}
\mathbf{M}_{ij}^{tt^\prime}=\delta^{t,t^\prime+1}
[(1-\mu)\delta_{ij}+\lambda \mathcal{A}_{ij}(t)].
\end{equation}
We mask the tensorial origin of the space through the map
$(i,t)\rightarrow\alpha=Nt+i$, where $1\leq\alpha\leq N\; t_{\rm max}$.
Thus $\mathbf{M}$ can be rewritten
\begin{equation}\label{tensor2}
\mathbf{M}=\left(
             \begin{array}{ccccc}
               0 & 1-\mu+\lambda \mathcal{A}(1) & 0 & \cdots & 0 \\
               0 & 0 & 1-\mu+\lambda \mathcal{A}(2) & \cdots & 0 \\
               \vdots & \vdots & \vdots & \vdots & \vdots \\
               0 & 0 & 0 & \cdots & 1-\mu+\lambda \mathcal{A}(t_{\rm max}-1) \\
               1-\mu+\lambda \mathcal{A}(t_{\rm max}) & 0 & 0 & \cdots &  0 \\
             \end{array}
           \right).
\end{equation}
Inserting Eq.~(\ref{tensor2}) into (\ref{s_line}), we have
\begin{equation}\label{array}
\widehat{S}(\tau)=\mathbf{M}\widehat{S}(\tau-1),
\end{equation}
where $\widehat{S}(\tau)$ is the probability that a node is in the
spreader state at each time step $t$ during $[\tau t_{\rm max},
  (\tau+1)t_{\rm max}]$. Here $\widehat{S}(\tau)$ increases
exponentially if the largest eigenvalue of $\mathbf{M}$, denoted
$\Lambda_1$, is larger than $1$. Thus the misinformation spreads, and
the threshold condition is \cite{valdano2015analytical}
\begin{equation}\label{condition}
\Lambda_1(\mathbf{M})=1,
\end{equation}
In an unweighted undirected network $\mathcal{G}$, the largest
eigenvalue $\Lambda_1(\mathbf{M})$ of $\mathbf{M}$ is
\begin{equation}\label{eig}
\Lambda_1(\mathbf{M})=\Lambda_1(\mathbf{P}^{1/t_{\rm max}}),
\end{equation}
where
\begin{equation}\label{P}
\mathbf{P}=\sum_{t=1}^{t_{\rm max}}(1-\mu+\lambda \mathcal{A}(t)).
\end{equation}

\section{Misinformation containing strategies} \label{con}

\noindent
Because misinformation spreading on social networks can induce social
instability, threaten political security, and endanger the economy, we
propose three strategies---random, targeted, and heuristic---for
containing the spread of misinformation in temporal networks using a
given fraction of {\it containing\/} nodes $f$. We first immunize a
fraction of $f$ nodes using a static containment strategy. The
misinformation then spreads on the residual temporal network. If node
$i$ is ``immunized,'' it cannot be infected (transmit) by the
misinformation received from neighbors (the misinformation to
neighbors). Mathematically the immunized node set is $\mathcal{V}$, and
the number of immunized nodes equals the number of elements in
$\mathcal{V}$, i.e., $|\mathcal{V}|= \lceil f\;N\rceil$. We set $v_i=1$
if node $i$ is immunized, otherwise $v_i=0$. After immunization,
Eqs.~(\ref{pit})--(\ref{s}) can be written
\begin{equation}\label{pit1}
p_i\left(t\right) = 1-\prod_j{\left[1-\lambda (1-v_j) \mathcal{A}_{ij}
\left(t\right) I_j\left(t\right) \right]},
\end{equation}
\begin{equation}\label{i1}
I_i(t+1)=I_i(t)-I_i(t)(1-v_i)p_i(t),
\end{equation}
and
\begin{equation}\label{s1}
S_i(t+1)=S_i(t)+I_i(t)(1-v_i)p_i(t)-\mu_i(t)S_i(t),
\end{equation}
respectively. In an effective containing strategy the misinformation
spreading dynamics is suppressed for a given fixed fraction $f$ of
immunized nodes, i.e., the objective function is
\begin{equation}\label{objective}
\min_{\mathcal{V}} {\lim_{t \rightarrow \infty}{\frac{1}{N} \sum_i R_i(t) }},
\end{equation}
where the constraint conditions are Eqs.~(\ref{i1})--(\ref{s1}) and
$V_i\,\, \in \left\{0,1\right\}$, $S_i \left(t\right) \in \,\left
[0,1\right ]$, $I_i\,\left(t\right) \in \,\left [0,1\right ]$, and $R_i
\left(t\right) \in \,\left [0,1\right ]$. Because the problem is
NP-hard, finding an accurate solution for large-scale temporal networks
and a finite immunization size $f$ is difficult. To address this
problem we propose three containment strategies.

\begin{itemize}

\item Strategy I: Random containment (RC).  The most used strategy for
  containing the spread of misinformation is randomly immunizing a
  fraction of $f$ nodes \cite{zuzek2015epidemic}.

\item Strategy II: Targeted containment (TC).  Another intuitive way is
  to immunize the nodes with highest average degree $\langle k\rangle$
  in the temporal network $\mathcal{G}$. Specifically, we first compute
  the average degree of each node $i$ as $\langle
  k_i\rangle=\frac{1}{t_{\rm max}}\sum_{t=1}^{t_{\rm max}}\sum_{j=1}^N
  \mathcal{A}_{ij}(t)$. We then rank all nodes in descending order in
  the vector $\mathcal{W}$ according to the average degree of each
  node. Finally we immunize the top $\lceil f N\rceil$ nodes of
  $\mathcal{W}$.

\item Strategy III: Heuristic containment (HC). Using the TC, we apply
  an HC strategy. Because TC is much better than RC, we perform the HC
  strategy by replacing the immunization nodes. When the repeat time is
  very large, the immunized nodes reach an optimal value.

\end{itemize}

\begin{description}

\item[(i)] We initialize a vector $\mathcal{W}$ according to the
  descending order of the average degree of nodes. The first $\lceil
  f\;N\rceil$ nodes of $\mathcal{W}$ are immunized, the final
  misinformation outbreak size is $R_o$, and $\mathcal{W}_0$ is denoted
  a set. The remaining nodes $\mathcal{W}_1
  =\mathcal{W}\backslash\mathcal{W}_0$ are denoted a set.

\item[(ii)] We randomly select nodes in $\mathcal{W}_0$ and
  $\mathcal{W}_1$, denoted $v_0$ and $v_1$, respectively. We switch
  their order in vector $\mathcal{W}$ and denote the new vector
  $\mathcal{W}_n$. We immunize the first $\lceil f\;N\rceil$ nodes
  $\mathcal{W}_n$ and compute the final misinformation outbreak size
  $R_n$.

\item[(iii)] When $R_n>R_o$, we update vector $\mathcal{W}$, i.e.,
  $\mathcal{W} \rightarrow \mathcal{W}_n$. Otherwise, there is no
  change.

\item[(iv)] We repeat steps (ii) and (iii) until
  $\frac{1}{t_s}\sum_{i=1}^{t_s}|\mathcal{W}-
  \mathcal{W}_n|<\epsilon^\prime$. In the simulations we set $t_s=100$ and
  $\epsilon^\prime=N^{-1}$.

\end{description}

\section{Numerical simulations} \label{num}

\noindent
For the activity-driven network, we set $N=10^3$, $t_{\rm max}=20$,
$\eta=10$, $m=50$, $\gamma=2.1$, and $\epsilon=10^{-3}$. For real-world
networks, we use the data collected by the Sociopatterns group
\cite{isella2011s}, which records the interactions among the
participants at a conference. The time resolution of the signal is 20
sec. Because the temporal network is sparse, it is difficult for the
information to spread in the original network. We thus aggregate the
temporal network using four windows, $w=30 {\rm min}$, $60 {\rm min}$,
$120 {\rm min}$, and $240 {\rm min}$. We average all simulation results
more than 1000 times.

We use variability to locate the numerical network-sized dependent
outbreak threshold \cite{shu2015numerical,shu2016recovery},
\begin{equation} \label{chi}
\chi=\frac{(\langle R^2\rangle-\langle R\rangle^2)^{1/2}}
{\langle R\rangle},
\end{equation}
where $R$ is the relative size of misinformation spreading at the steady
state.  At the outbreak threshold $\lambda_c$, $\chi$ exhibits a peak.
When $\lambda\leq\lambda_c$, the global misinformation does not break
out, but when $\lambda>\lambda_c$ the global misinformation does break
out.

\begin{figure}
\begin{center}
\epsfig{file=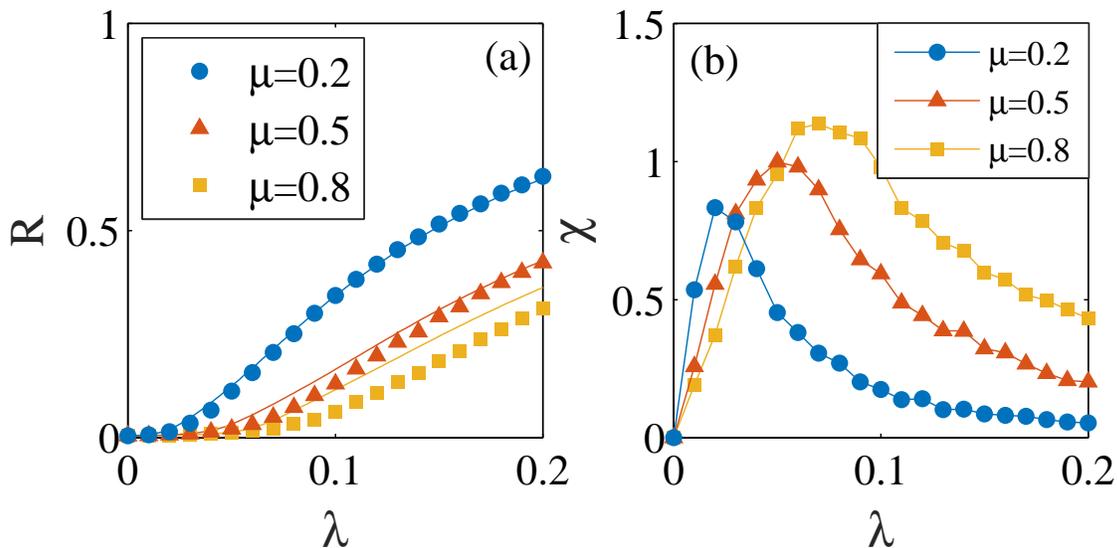,width=1\linewidth}
\caption{Misinformation spreading on activity-driven temporal networks.
  (a) The final misinformation outbreak size $R$, and the numerical
  variability $\chi$ versus the transmission probability $\lambda$ with
  different values of recovery probability $\mu$. In (a), the symbols
  are the numerical simulation results, and lines are the theoretical
  predictions.  }\label{figure1}
\end{center}
\end{figure}

\begin{figure}
\begin{center}
\epsfig{file=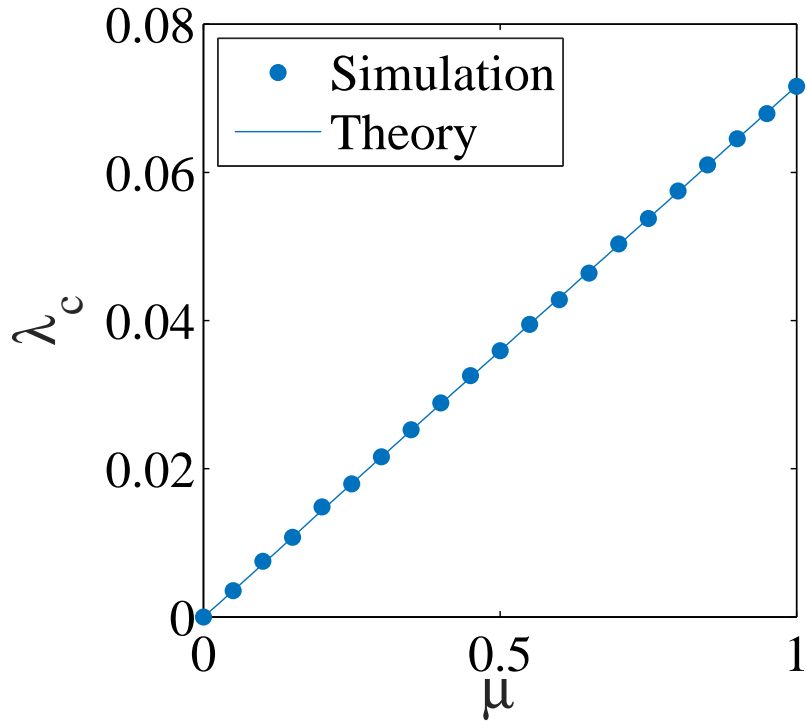,width=1\linewidth}
\caption{The outbreak threshold $\lambda_c$ versus $\mu$.  The symbols
  represent the stochastic simulation threshold, and line is the
  theoretical prediction.  }\label{figure2}
\end{center}
\end{figure}

Figure~\ref{figure1} shows the misinformation spreading on
activity-driven networks. Note that the final misinformation outbreak
size $R$ increases with $\lambda$. The larger the recovery probability
$\mu$, the lower the values of $R$ because spreader nodes are less
likely to transmit the misinformation to stifler neighbors [see
  Fig.~\ref{figure1}(a)].  Note that our theoretical and numerical
predictions of the final misinformation outbreak size $R$ agree.
Figure~\ref{figure1}(b) shows the variability $\chi$ as a function of
$\lambda$. There is a peak at the misinformation outbreak threshold
$\lambda_c$. Figure~\ref{figure2} shows $\lambda_c$ versus $\mu$ in
which $\lambda_c$ increases linearly with $\mu$. The theoretical
predictions of $\lambda_c$ obtained from Eq.~(\ref{condition}) agree
with the stochastic simulations.
	
\begin{figure}
\begin{center}
\epsfig{file=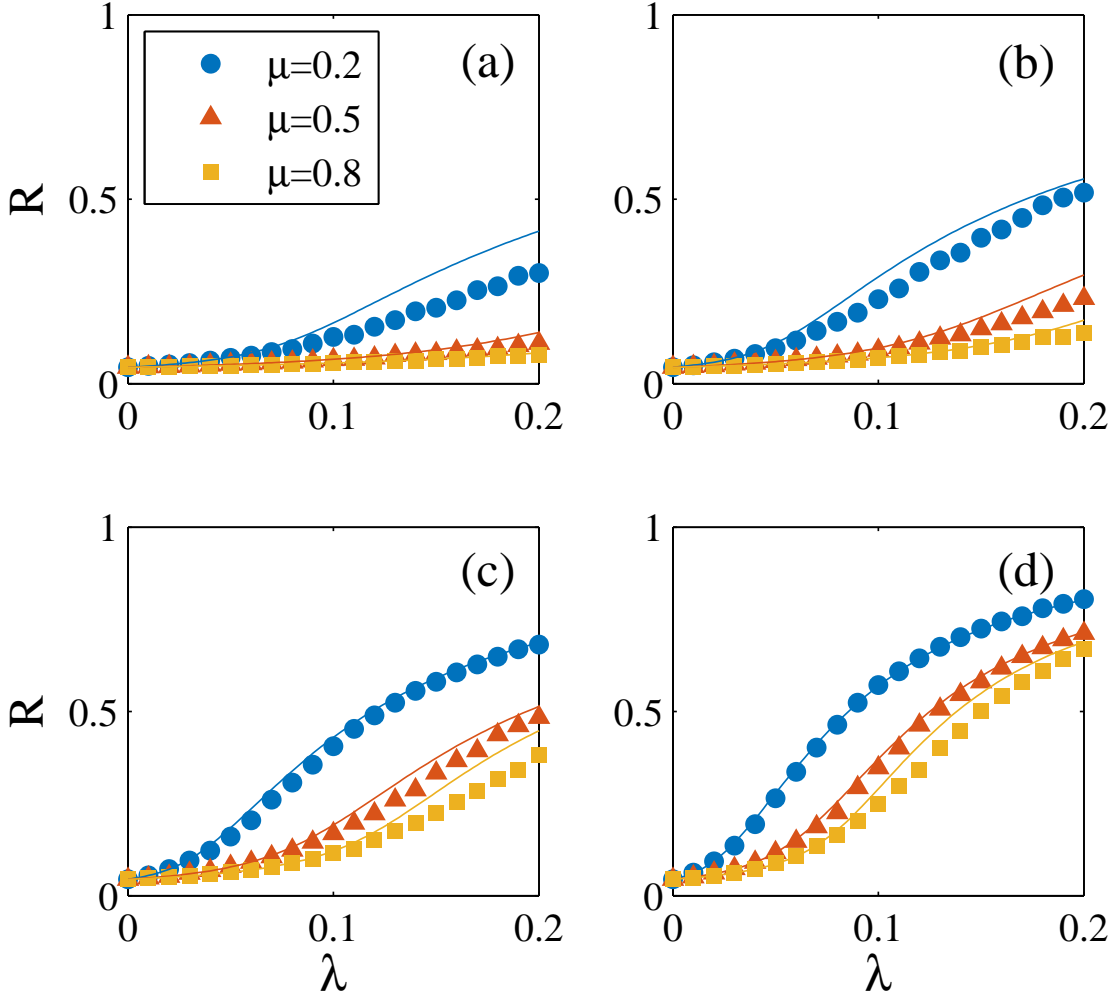,width=1\linewidth}
\caption{Misinformation spreading on real-world temporal networks.  The
  final misinformation outbreak size $R$ versus $\lambda$ on ACM
  Hypertex conference data set with aggregating window (a) $w=30~{\rm
    min}$, (b) $w=60~{\rm min}$, (c) $w=120~{\rm min}$ and (d)
  $w=240~{\rm min}$. The symbols are the numerical simulation results,
  and lines are the theoretical predictions.  }
\label{figure3}
\end{center}
\end{figure}

Figure~\ref{figure3} shows misinformation spreading in real-world
temporal networks. As in Fig.~\ref{figure1}, $R$ increases with
$\lambda$ and decreases with $\mu$.  As in SIR epidemic spreading
\cite{ogura2017optimal}, the effective outbreak threshold
$(\lambda/\mu)_c$ is a constant value. In addition, when the value
aggregating window $w$ is small, there are fewer opportunities for
spreaders to transmit the misinformation to stifler neighbors, thus the
misinformation does not break out globally, i.e., there are smaller
values of $R$ for smaller $w$. Once again our theoretical results agree
with the numerical simulations.
	
\begin{figure}
\begin{center}
\epsfig{file=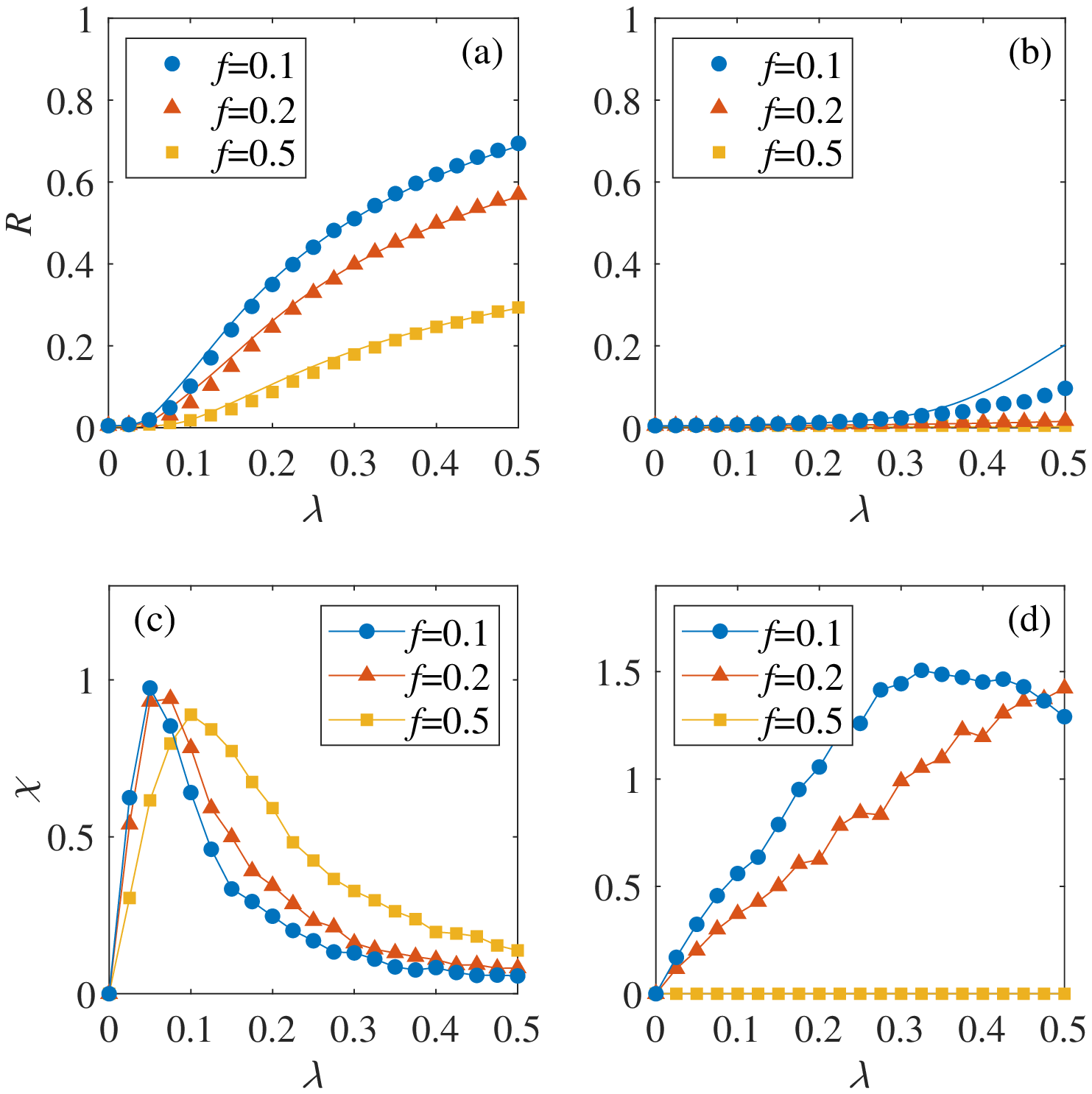,width=1\linewidth}
\caption{Containing misinformation spreading dynamics on activity-driven
  temporal networks. The final misinformation outbreak size $R$ (a) and
  variability $\chi$ (c) versus the transmission probability $\lambda$
  under the RC strategy. $R$ (b) and $\chi$ (d) versus $\lambda$ under
  the TC strategy. In (a) and (b), the lines are the theoretical
  predictions, and symbols are the numerical simulation results.
}\label{figure4}
\end{center}
\end{figure}

We next examine the performances of our proposed strategies for
mitigating misinformation spreading on artificial and real-world
temporal networks. Figure~\ref{figure4} shows $R$ versus $\lambda$ for
different values of the fraction of containing nodes $f$. Note that $R$
decreases with $f$ because no more nodes receive the
misinformation. Note also that the TC strategy performs much better than
the RC strategy because the higher degree nodes $\langle k\rangle$ are
contained, and spreaders can no longer transmit the misinformation to
stiflers. Thus when we immunize the same fraction of containing nodes,
the values of $R$ for the TC strategy are smaller than those for the RC
strategy. For example, when $f=0.5$, using the RC strategy $\approx 30
\%$ of the nodes are informed by the misinformation, but using the TC
strategy none are informed by the misinformation. In addition, the
outbreak threshold $\lambda_c$ decreases with
$f$. Figures~\ref{figure4}(c)--\ref{figure4}(d) show the numerical
predictions for $\lambda_c$.

\begin{figure}
\begin{center}
\epsfig{file=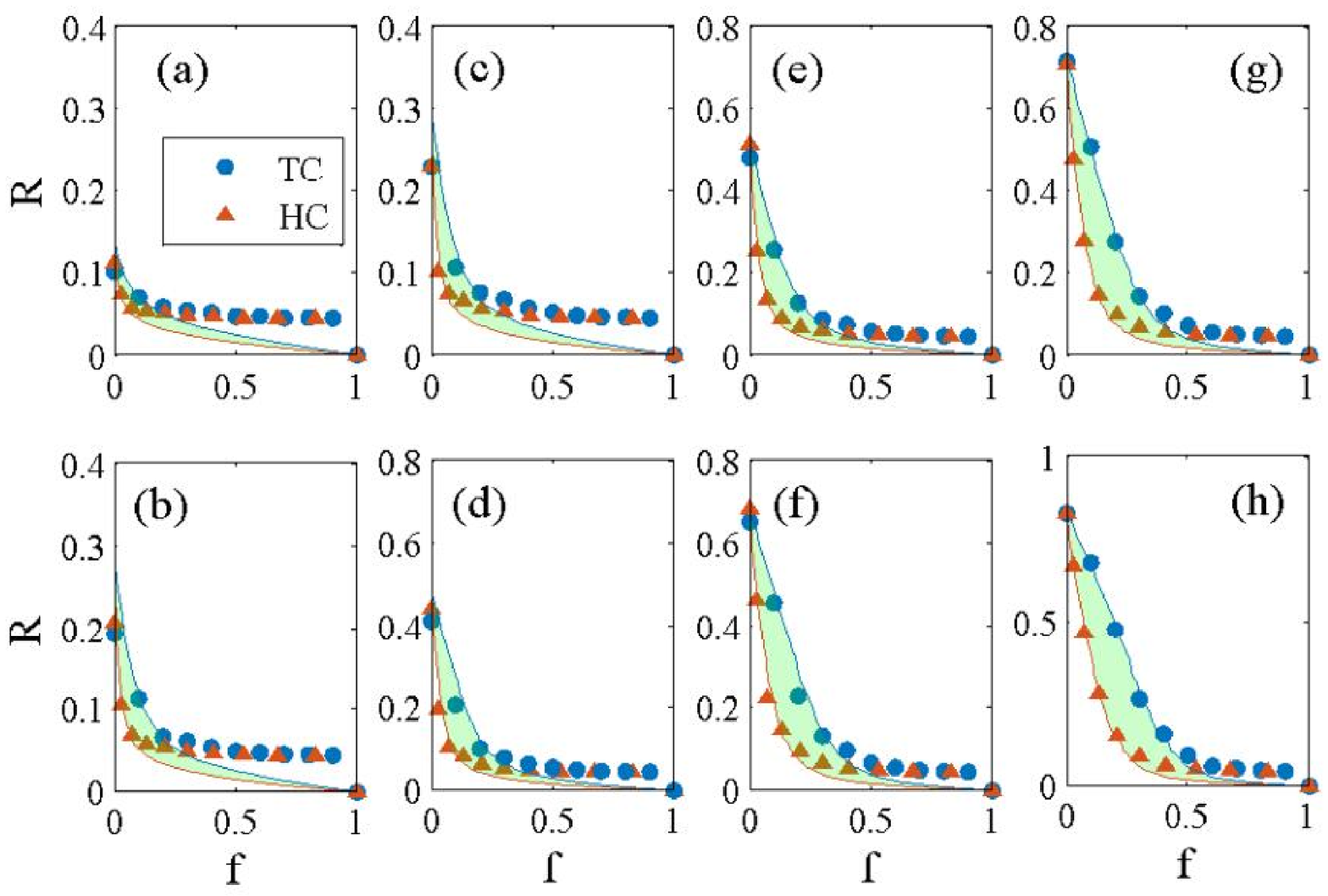,width=1\linewidth}
\caption{Effectiveness of TC and HC strategies on activity-driven
  networks. The final misinformation outbreak size $R$ versus $f$ for a
  given values of $\lambda=0.2$ (a) and $\lambda=0.3$ (b). The lines and
  symbols are the theoretical and the numerical predictions of $R$,
  respectively.  }\label{figure7}
\end{center}
\end{figure}

Figure~\ref{figure7} shows that the HC strategy performs better than the
TC strategy. We set To this end, we set $\lambda=0.2$ and 0.3 in
\ref{figure7}(a) and \ref{figure7}(b), respectively, and compare the
final misinformation spreading size $R$ with $f$. We find that the
misinformation spreading dynamics are suppressed when we use the HC
strategy. For example, when $f=0.3$ and $\lambda=0.3$, few nodes are
informed by the misinformation, i.e., $R\approx0.01$ when we use the HC
strategy, but when we use the TC strategy to contain the misinformation
the fraction of nodes informed by the misinformation is finite, i.e.,
$R\approx 0.25$. Our theoretical results agree with the numerical
simulation results.

\begin{figure}
\begin{center}
\epsfig{file=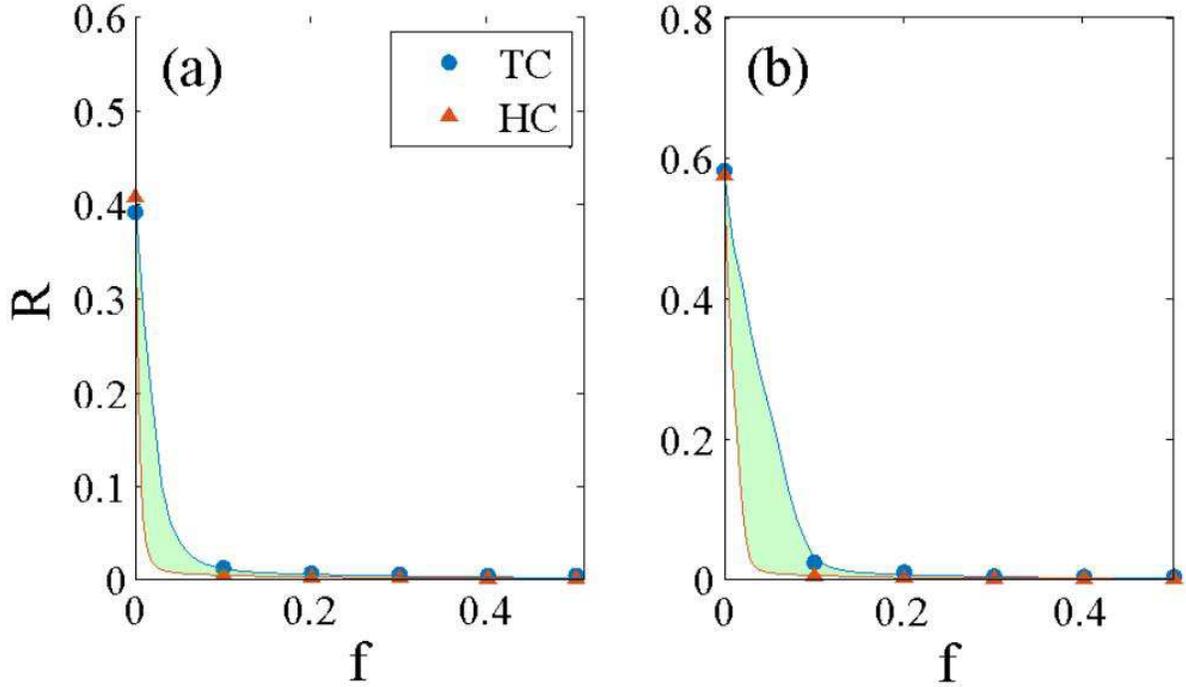,width=1\linewidth}
\caption{The effective outbreak threshold $(\lambda/\mu)_c$ versus the
  fraction of containing nodes $f$ on activity-driven networks for RC
  strategy (a), TC strategy (b), HC strategy (c) and the comparision of
  the three strategies (d). The lines and symbols are the theoretical
  and numerical predictions of $(\lambda/\mu)_c$, respectively. The
  vertical line represents the critical probability $f_c$.
}\label{figure5}
\end{center}
\end{figure}

An effective containing strategy with a fraction of immunized node $f$
and a small outbreak threshold $\lambda_c$ greatly decreases the final
misinformation outbreak size $R$.  Figure~\ref{figure5} shows the
effective outbreak threshold $(\lambda/\mu)_c$ versus $f$ on
activity-driven networks for the RC, TC, and HC strategies. Here
$(\lambda/\mu)_c$ increases with $f$, and $(\lambda/\mu)_c$ is the
largest using the HC strategy when $f$ is fixed. When $f$ is
sufficiently large, no $\lambda$ value can induce a global
misinformation outbreak.  We denote $f_c$ the critical probability that
at least a fraction of $f_c$ nodes must be containing to halt
misinformation speading in temporal networks. We find that the values of
$f_c$ for the HC containing strategy are the smallest of all containing
strategies. The $f_c$ value for the RC strategy is 5 times the $f_c$
value for the TC strategy, and the $f_c$ value for the TC strategy is
2.5 times the $f_c$ value for HC strategy.  Thus the HC strategy is the
most effective.

\begin{figure*}
\begin{center}
\epsfig{file=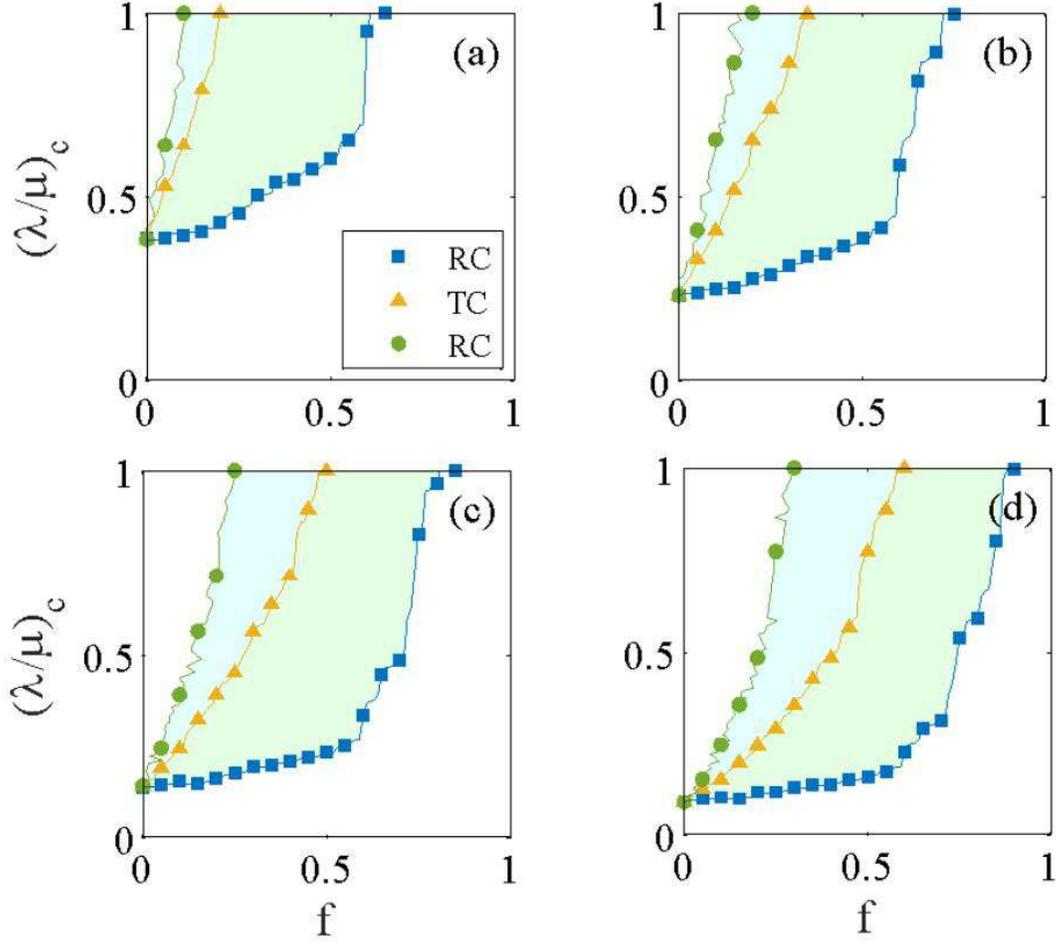,width=0.9\linewidth}
\caption{Effectiveness of TC and HC strategies on ACM Hypertex
  conference data set. $R$ versus $f$ with aggregating window (a)
  $w=30~{\rm min}$, (c) $w=60~{\rm min}$, (e) $w=120~{\rm min}$ and (g)
  $w=240~{\rm min}$ for $\lambda=0.2$. $R$ versus $f$ with (b)
  $w=30~{\rm min}$, (d) $w=60~{\rm min}$, (f) $w=120~{\rm min}$ and (h)
  $w=240~{\rm min}$ for $\lambda=0.3$. The lines and symbols are the
  theoretical and numerical predictions of $R$, respectively.
}\label{figure8}
\end{center}
\end{figure*}

\begin{figure}
\begin{center}
\epsfig{file=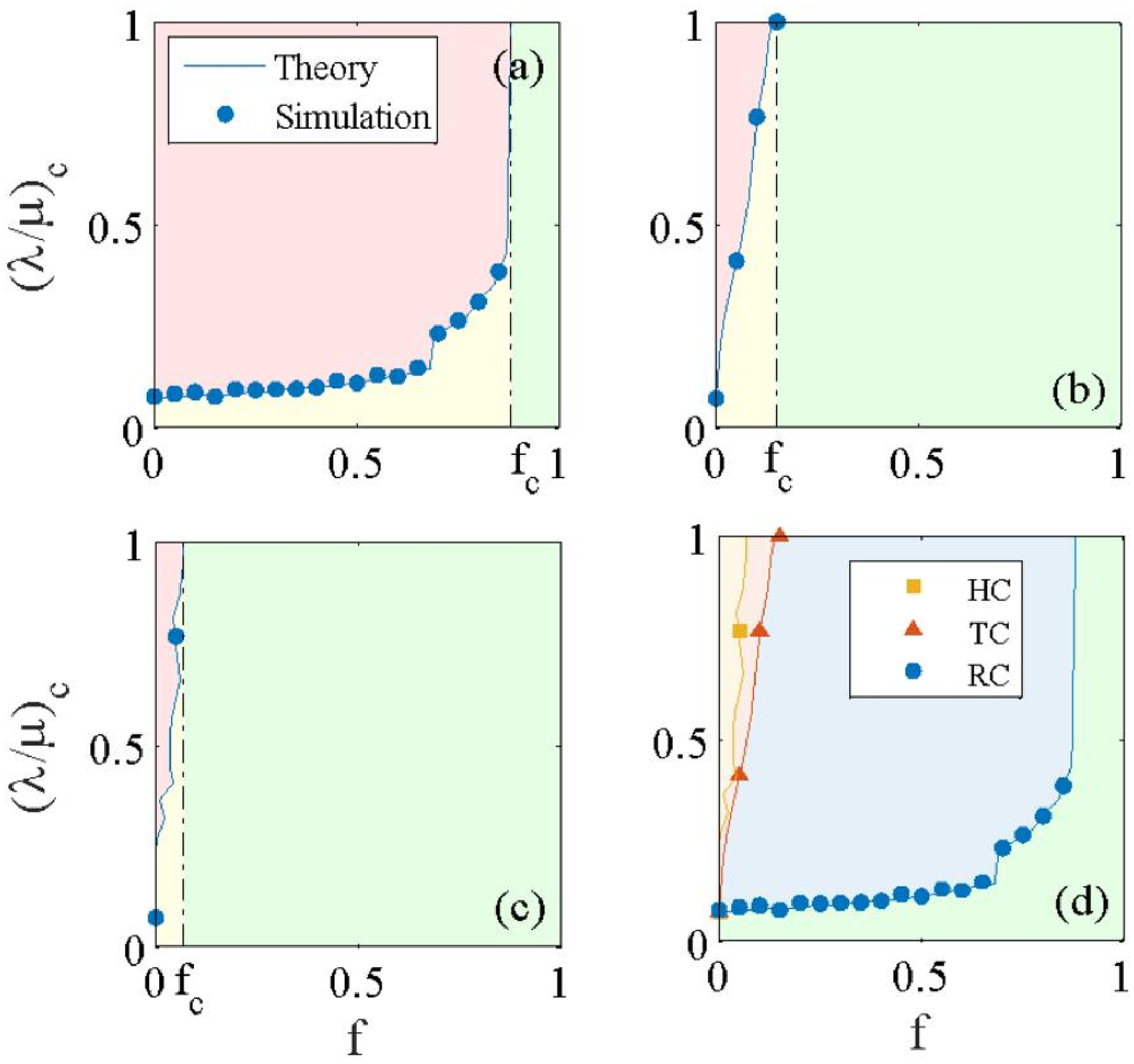,width=1\linewidth}
\caption{Comparing the effectiveness of RC, TC and HC strategies on ACM
  Hypertex conference data set with aggregating window (a) $w=30~{\rm
    min}$, (b) $w=60~{\rm min}$, (c) $w=120~{\rm min}$ and (d)
  $w=240~{\rm min}$. The lines and symbols are the theoretical and
  numerical predictions of $(\lambda/\mu)_c$, respectively.
}\label{figure6}
\end{center}
\end{figure}

We finally examine real-world networks to varify the effectiveness of
our proposed three strategies. Figure~\ref{figure8} compares the
performances of the TC and HC strategies by examining $R$ versus $f$ for
given values of $\lambda$. As in Fig.~\ref{figure7}, the HC strategy
most effectively contains the misinformation spreading on temporal
networks irrespective of the values of $\lambda$. In addition,
Fig.~\ref{figure6} shows that the effective outbreak threshold
$(\lambda/\mu)_c$ is the smallest when using the HC strategy. Thus our
theory accurately predicts the numerical simulation results.

\section{Conclusions} \label{conc}

\noindent
We have systematically examined the dynamics of misinformation spreading
on temporal networks. We use activity driven networks to describe
temporal networks, and use a discrete Markovian chain to describe the
spreading dynamics. We find that the global misinformation outbreak
threshold correlates with the topology of temporal networks. Using
extensive numerical simulations, we find that our theoretical
predictions agree with numerical predictions in both artificial and
real-world networks.

To contain misinformation spreading on temporal networks, we propose
three strategies, random containing (RC), targeted containing (TC), and
heuristic containing (HC) strategies. We perform numerical simulations
and a theoretical analysis on both artificial and four real-world
networks and find that the HC strategy outperforms the other two
strategies, maximizes the outbreak threshold, and minimizes the final
outbreak size. Our proposed containing strategy expands our
understanding of how to contain public sentiment and maintain social
stability.

\acknowledgments

\noindent
This work was partially supported by the China Postdoctoral Science
Foundation (Grant No.~2018M631073), and Fundamental
Research Funds for the Central Universities.
LAB thanks UNMdP and CONICET (PIP 00443/2014) for financial support.

%

\end{document}